# Model of information diffusion

D.V. Lande,
ElVisti Information Center, Ukraine, Kiev

The system of cellular automata, which expresses the process of dissemination and publication of the news among separate information resources, has been described. A bell-shaped dependence of news diffusion on internet-sources (web-sites) coheres well with a real behavior of thematic data flows, and at local time spans - with noted models, e.g., exponential and logistic ones.

***Key words***: *information flows, dissemination of information, cellular automata, model of information diffusion*

At present, when scope and dynamics of the information publication in the Internet give all grounds to use the term information flows [1], to study their dynamics becomes urgent. The diversity in the behavior of publications according to various topics and the complexity of cross-impact of various publications prove the necessity to seek new methods which have not been known in this field before. Probably, at this stage in the sphere of modeling complicated information processes, a success can be made through the synthesis of quite simple algorithms and concepts.

The theory of cellular automata, suggested by J. von Neiman over thirty years ago [2] and developed by S. Wolfram in a fundamental monograph [3] can be recognized as a promising one in this field.

A cellular automaton is a discrete dynamic system, aggregate of similar cells connected with each other. All cells form a net of cellular automata. The condition of each cell is determined by that of the cells which are in its local vicinity/neighborhood. A set of its "nearest neighbors" is called the neighborhood of a finite automation with number *j*. The condition of *j*-cellular automation in the time moment *t + 1* is defined in the following way:

$$y_j(t+1) = F(y_j(t), O(j), t), \qquad (1)$$

where *F* – a certain rule which can be expressed, e.g., in the language of Boolean algebra, *O(j)* – neighborhood, *t* – a step. Cellular automata correspond to such rules:
- the change of the meanings of all cells occur simultaneously (unit of time measurement is a step);
- a net of cellular automata is homogenous, i.e., the rules of the condition changing is the same for all cells;
- a cell can be influenced only by the cells from its local neighborhood;
- a set of cell conditions is final.

The theory of cellular automata is known to be used to analyze innovation diffusion, the process being very similar to news dissemination in the Internet [4]. A similar model functions according to the following rules: an individual capable of receiving innovation corresponds to one cell, which may be in two conditions: 1 – news accepted; 0 – news not accepted. It is assumed that automation having received the innovation once, remembers it for ever (condition 1 – cannot be changed). Automation makes a decision to receive news, being oriented on the opinion of eight closest neighbors, i.e., if in the neighborhood of a given cell there are *m* supporters of the news, and *p*- probability of receiving the news (being generated in the course of the model work),
then if $pm > R$,
where *R* – a fixed threshold meaning, $\qquad (2)$
the cell receives innovation (receives meaning 1).

In addition, the dynamics of information diffusion has some additional properties, considered by the author. Under the same conditions a cell is suggested to be in one of the three conditions: 1 - "fresh news" (a cell is in black color); 2 - a little bit obsolete news, retained as information (a grey cell); 3 - a cell does not have information transmitted by the news message (a cell is white; information did not reach or was forgotten). The rules of news dissemination are the following:

- at the beginning the whole field consists of white cells except for one – a black cell which was the first "to receive" the news (Fig. 1a);
- a white cell can change its color into black or remain white (it may receive the news or remain "in ignorance");
- a white cell changes its color, if the condition analogical to (2) is met in a model of innovation diffusion: $pm > 1$ ($m$ – the number of black cells, but if $m<3$, the meaning of $p$ increases by 1.5 times);
- if a cell is black, and cells around it are exclusively black and grey, it changes its color into grey ( the news gets obsolete and remains as the message);
- if a cell is grey, and cells around it are exclusively grey and black, it changes its color into white (well-known information is being forgotten).

A described-above system of cellular automata expresses the process of dissemination and publication of the news among other information sources quite realistically. In a 40 x 40 field (the authors chose dimensions exclusively for the sake of better demonstration) the condition of the cellular automata system is completely stabilized within a limited quantity of steps, i.e., the evolution process is convergent. The example of the model work is given in Fig. 1.

Numerous experiments with a given cellular automation, available in the Internet at the address http://edu.infostream.ua/newsk.pl, show that a period of its convergence ranges form 80 to 150 steps. Typical dependences of the cell number (sequences of a duplicate cell number), which are in various conditions depending on a step number, are presented in Fig. 2. Analyzing the figures presented in this paper, one should pay attention to such peculiarities: 1 - a total number of cells which are in all three conditions at each step of iteration constantly and equally to the field size, 2 – while stabilizing cellular automata, relation of grey, white and black cells is about 0.75 : 0.25 : 0; there exists a cross-point of curves determined by all three sequences at the level of 33% each.

Special attention should be paid to the dependence formed by black cells. A form/view of a given curve agrees with "life" dynamics of the news, first it is actively diffused, covering all new corners of information space, then certain saturation occurs, and it is no longer new for most of the recipients, it becomes simple the information and then it is forgotten.

A detailed analysis of the dependences received made it possible to draw analogies between a given model of "information diffusion" and further analytical considerations. The results of modeling allow us to presume that the evolution of grey cells is described with a continuous function:

$$x_g = f\left(t, \tau_g, \gamma_g\right), \qquad (3)$$

where $t$ - time (evolution step), $\tau_g$ - shift in time, ensuring the receipt of required fragment of an analytical function, $\gamma_g$ - a parameter of a curve slope of a given function.

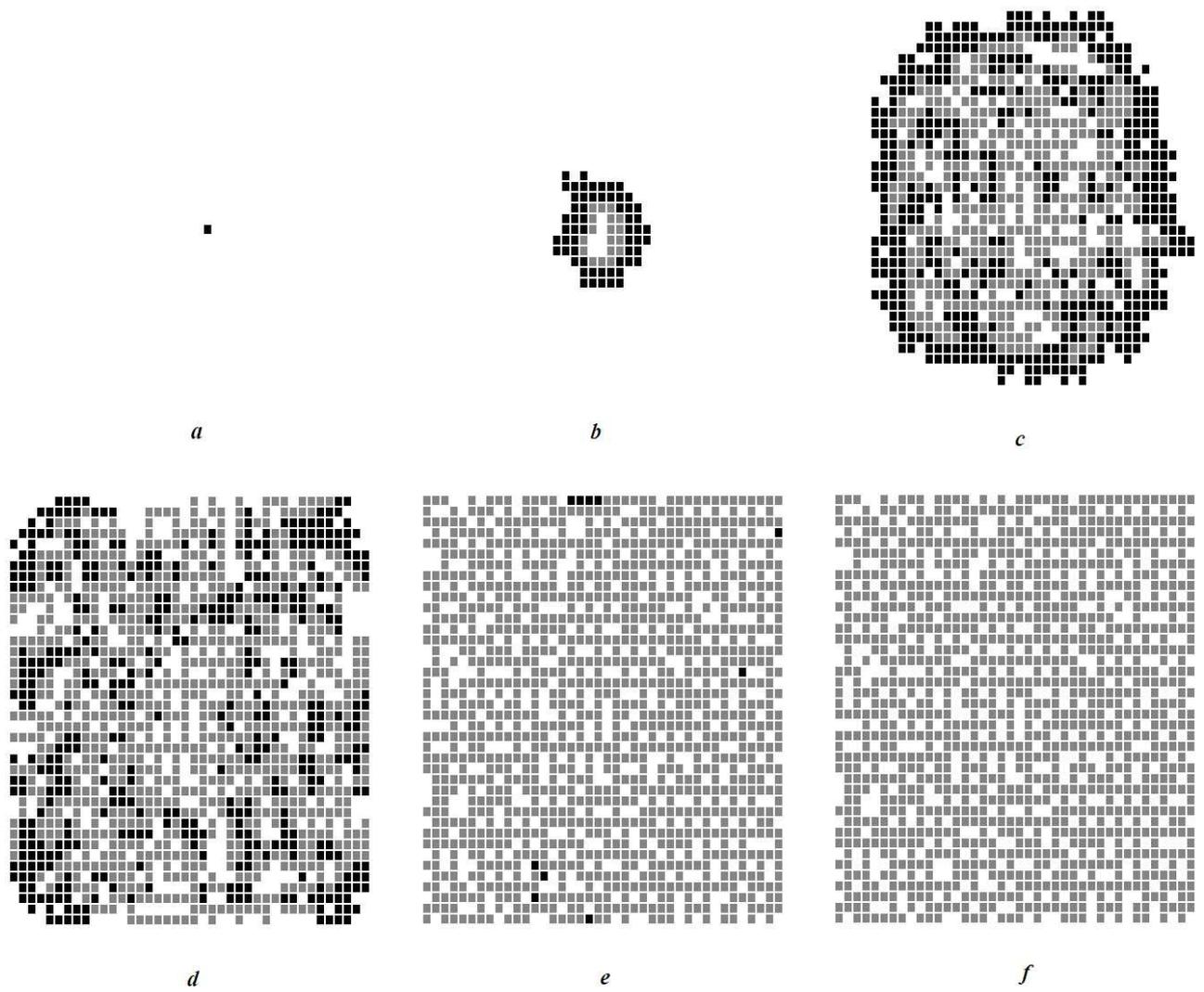

*Fig.1. The evolution process of the cellular automata system of "news diffusion": a – initial condition; b-e – intermediate condition; f – finite condition*

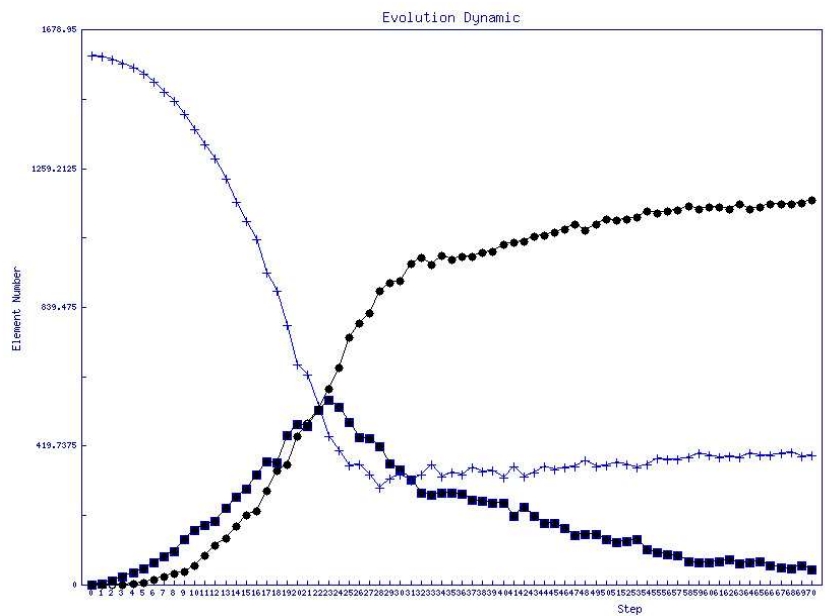

*Fig. 2. The number of cells of each color depending on an evolution step: white cells - (✚); grey cells - (●); black cells – (■)*

Correspondingly, the dynamics of white cells $x_w$ (the number of cells at the moment $t$) can be modeled with an "inverted" function $x_g$ with its analogous parameters:

$$x_w = 1 - f(t, \tau_w, \gamma_w). \qquad (4)$$

It has been mentioned above, as the balance condition is always met, i.e., the total number of cells at a particular moment is always constant, the condition of standardizing can be written down:

$$x_g + x_w + x_b = 1, \qquad (5)$$

where $x_w$ - the number of black cells at the time moment $t$.

Thus, according to (5):

$$x_b = 1 - x_g - x_w = f(t, \tau_w, \gamma_w) - f(t, \tau_g, \gamma_g). \qquad (6)$$

A view of the dependence, given in Fig. 2, allows presuming that the following expression can be chosen as function $f(t, \tau, \gamma)$:

$$f(t, \tau, \gamma) = \frac{C}{1 + e^{\gamma(t-\tau)}}, \qquad (7)$$

where $C$ - some normalizing constant.

In fig. 3 a plot of dependences $x_g, x_w, x_b$ on the evolution step of the cellular automata system, received as a result of analytical modeling, was expressed with formulae:

$$x_g = \frac{0.75}{1 + e^{-0.15(t-30)}};$$

$$x_w = 1 - \frac{0.75}{1 + e^{-0.25(t-20)}}; \qquad (8)$$

$$x_b = 0.75 \left( \frac{1}{1 + e^{-0.25(t-20)}} - \frac{1}{1 + e^{-0.15(t-30)}} \right).$$

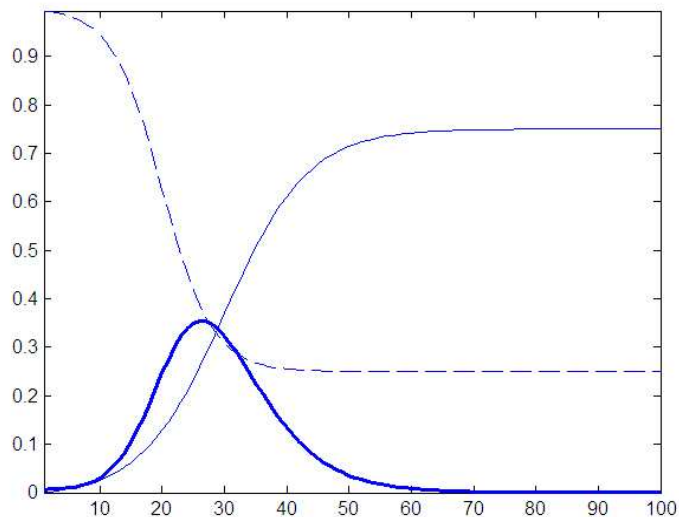

*Fig.3. Continuous dependences, received as a result of analytical modeling, depending on an evolution step: full line - grey ($x_g$); dotted line – white ($x_w$); full heavily drawn line – black ($x_b$)*

As a conclusion, it is worth mentioning that a received bell-shaped dependence of news diffusion on Internet-sources (web-sites) agrees well with a real behavior of thematic information flows, and on local time spans – with noted models, e.g., exponential and logistic ones. The comparison of experimental dependences with the results of analytical modeling proves high accuracy of the experimental data by model curves.